\documentclass[prl,aps,twocolumn,showpacs,superscriptaddress]{revtex4}
\usepackage{graphicx}
\usepackage{ifpdf}
\ifpdf
\usepackage[
	pdfpagemode=None,
  colorlinks=true,
	linkcolor=black,
	filecolor=black,
	urlcolor=black,
	citecolor=black,
	bookmarks=false,
	pdftex=true,
	plainpages=false,
	hypertexnames=false,
	pdfpagelabels=true,
	hyperindex=true]{hyperref}
\DeclareGraphicsExtensions{.pdf}
\else
\DeclareGraphicsExtensions{.eps}
\fi

\usepackage{verbatim}
\begin{document}

\title{Precision Measurement of $^{11}$Li moments: influence of halo neutrons on the $^9$Li core}

\author{R.~Neugart}
\affiliation{Institut f\"ur Physik, Universit\"at Mainz, D-55099 Mainz, Germany}
\author{D.~L.~Balabanski}
\altaffiliation[Present address: ] {INRNE, Bulgarian Academy of
Sciences, BG-1784 Sofia, Bulgaria.}
\affiliation{Instituut voor
Kern- en Stralingsfysica, K.U.Leuven, B-3001~Leuven, Belgium}
\author{K.~Blaum}
\altaffiliation[Present address: ] {Max-Planck-Institut f\"ur
Kernphysik, Saupfercheckweg 1, D-69177 Heidelberg, Germany.}
\affiliation{Institut f\"ur Physik, Universit\"at Mainz, D-55099
Mainz, Germany} \affiliation{Physics Department, CERN, CH-1211
Geneva~23, Switzerland}
\author{D.~Borremans}
\affiliation{Instituut voor Kern- en Stralingsfysica, K.U.Leuven, B-3001~Leuven, Belgium}
\author{P.~Himpe}
\affiliation{Instituut voor Kern- en Stralingsfysica, K.U.Leuven, B-3001~Leuven, Belgium}
\author{M.~Kowalska}
\affiliation{Institut f\"ur Physik, Universit\"at Mainz, D-55099 Mainz, Germany}
\affiliation{Physics Department, CERN, CH-1211 Geneva~23, Switzerland}
\author{P.~Lievens}
\affiliation{Laboratorium voor Vaste-Stoffysica en Magnetisme, K.U.Leuven, B-3001~Leuven, Belgium}
\author{S.~Mallion}
\affiliation{Instituut voor Kern- en Stralingsfysica, K.U.Leuven, B-3001~Leuven, Belgium}
\author{G.~Neyens}
\affiliation{Instituut voor Kern- en Stralingsfysica, K.U.Leuven, B-3001~Leuven, Belgium}
\author{N.~Vermeulen}
\affiliation{Instituut voor Kern- en Stralingsfysica, K.U.Leuven, B-3001~Leuven, Belgium}
\author{D.~T.~Yordanov}
\affiliation{Instituut voor Kern- en Stralingsfysica, K.U.Leuven, B-3001~Leuven, Belgium}


\date{\today}


\begin{abstract}
The electric quadrupole moment and the magnetic moment of the
$^{11}$Li halo nucleus have been measured with more than an order
of magnitude higher precision than before, $\vert Q
\vert=33.3(5)$~mb and $\mu=3.6712(3)\,\mu_N$, revealing a
8.8(1.5)$\%$ increase of the quadrupole moment relative to that of
$^9$Li. This result is compared to various models that aim at
describing the halo properties.  In the shell model an increased
quadrupole moment points to a significant occupation of the $1d$
orbits, whereas in a simple halo picture this can be explained by
relating the quadrupole moments of the proton distribution to the
charge radii. Advanced models so far fail to reproduce
simultaneously the trends observed in the radii and quadrupole
moments of the lithium isotopes.
\end{abstract}
\maketitle

Since nuclear physicists could produce and investigate bound
systems of nucleons in many possible combinations, a wealth of
isotopes with unexpected properties have been discovered.  For
example, some neutron-rich isotopes of light elements, such as
$^{11}$Li, were found to have exceptionally large radii
\cite{Tan96}. Upon discovery in 1985, this phenomenon was
attributed to either large deformation or to a long tail in the
matter distribution \cite{Tan85}. Deformation was soon excluded by
the spin and magnetic moment of $^{11}$Li belonging to a spherical
$\pi p_{3/2}$ state \cite{Arn87}. Considering the weak binding of
the last two neutrons
\cite{Thi75}, one could conclude that such a nuclear system
consists of a core with two loosely bound neutrons around it
\cite{Han87}. This is the concept of 'halo' nuclei which has been
related to similar phenomena in atomic and molecular physics
\cite{Jen04}, showing the universality of the concept. To fully
unravel the
mechanisms leading to the existence of halo nuclei, many types of
experiments have been devoted to the investigation of their
properties. An observable that gives information on the nuclear
charge deformation is the spectroscopic quadrupole moment. By
comparing the quadrupole moment of $^{11}$Li to that of $^{9}$Li,
one can investigate how the two halo neutrons modify the
deformation of the core which contains the three protons. Already
fifteen years ago, a first attempt to do so giving
$Q(^{11}$Li$)/Q(^9$Li$)=1.14(16)$, suggested just a slight
increase in agreement with the halo concept \cite{Arn92}.
Unbiasedly, for a nucleus with a neutron magic number of $N=8$ one
would expect a
minimum value of the quadrupole such as for
$^{13}$B \cite{Izu96}. If $^{11}$Li has a larger quadrupole moment
than $^9$Li, it can not be considered as semi-magic, and the two
halo neutrons have to be responsible for an expansion or
polarization of the proton distribution in the core. The latter,
in terms of the shell model, must be understood by an excitation
of halo neutrons to the $1d$ orbits, and a precise
value
of the $^{11}$Li quadrupole moment may provide evidence for this.
The effect of a more extended charge distribution can be estimated
on the basis of a recent laser spectroscopy measurement of the
charge radius \cite{San06}. The increase of the charge radius for
$^{11}$Li, as well as other properties of Li isotopes are well
described by cluster models \cite{Var02,Des97} which also
explain the large
quadrupole moment of $^7$Li
\cite{Var95}. For $^{11}$Li they predict a quadrupole moment over
30$\%$ larger than that of $^9$Li. Again, this calls for a more
precise measurement.

In this Letter, we report the measurement of the quadrupole moment
of $^{11}$Li relative to that of $^{9}$Li, intended to resolve a
difference at the percent level. Improvements over the first study
were made to gain an order of magnitude in precision. These
concern the experimental method which is based on the nuclear
magnetic resonance (NMR) technique. Spin-polarized beams of
$\beta$-decaying isotopes are implanted into a crystal with a
non-cubic lattice structure placed between the poles of an
electromagnet. Due to the electric field gradient $V_{zz}$ in such
a crystal, in combination with the static magnetic field $B$, the
Zeeman levels of the nuclear spin are shifted by the
$m_I$-dependent quadrupole interaction which is proportional to
the interaction constant $\nu_Q = eQV_{zz}/h$ (as illustrated e.g.
in \cite{Bor05} for $^9$Li) and thus to the nuclear quadrupole
moment $Q$. The frequency $\nu_{\rm rf}$ of an additionally
applied radio-frequency (rf) magnetic field is scanned over the
resonances between adjacent $m_I$ quantum states.
Whenever $\nu_{\rm rf}$ matches one of the transition frequencies,
a reduction of the spin polarization shows up as a drop in the
$\beta$-decay asymmetry \cite{Neu06}. The resonances are
equidistant and symmetric with respect to the Larmor frequency
$\nu_L$, with the spacing given by $\Delta_r= 6\nu_Q/[4I(2I-1)]$.

The resonance amplitude is enhanced by more than an order of
magnitude by applying all resonance frequencies simultaneously,
thus mixing all $m_I$ states \cite{Arn92}. For nuclei with spin
$I=3/2$, such as $^{9}$Li and $^{11}$Li, this
involves the three correlated frequencies $\nu_L$, $\nu_L+\Delta$
and $\nu_L-\Delta$. A spectrum is then measured as a function of
the parameter $\Delta$, with the resonance value $\Delta_r$
determining the quadrupole interaction frequency $\nu_Q$. The
signal enhancement and the width, thus also the error on $\nu_Q$,
depend on the sufficiently precise knowledge of the Larmor
frequency,
measured in a crystal with cubic lattice structure.

In the previous study of Arnold et al.\ \cite{Arn92} the resonance
width, mainly determined by
variations of the electric field gradient over the implantation
sites in
the LiNbO$_3$ crystal, was of the same order as the
splitting $\Delta_r$ itself. Now, we have reduced this width by an
order of magnitude
to less than 2 kHz \cite{Bor05}, by
using
a metallic Zn crystal. This also means that less rf power is
needed to saturate the resonances.
The known magnetic moment $\mu(^{11}$Li$) = 3.6673(25)\,\mu_N$
\cite{Arn87} corresponds to an uncertainty of 3.5 kHz on the
Larmor frequency of about 5 MHz, but an accuracy much better than
the expected line width is needed for a multiple-rf resonance
measurement. This is achieved by implanting into a Si crystal
\cite{Bor05} where the NMR line width is an order of magnitude
smaller than in Au \cite{Cor83} or LiF \cite{Arn87}.

The experiments have been performed at ISOLDE/ CERN.  Beams of
$^{9}$Li and $^{11}$Li were produced by a 1.4 GeV proton beam
($3\times 10^{13}$ protons per pulse every 2.4 s) on a thin-foil
Ta \cite{Ben02} or a conventional UC$_2$ target. With the short
release times of both targets typical production rates of a few
1000 ions/pulse were realized for the short-lived $^{11}$Li
($T_{1/2}=8.5(2)$ ms). The experimental setup and the method of
optically polarizing Li beams have been described in detail in
Borremans et al.\ \cite{Bor05}, where we report
results from our studies on the resonance properties in different
crystals for the less exotic isotopes $^{8}$Li and $^{9}$Li.

\begin{figure}
\centerline{\scalebox{0.65}[0.65]{\includegraphics{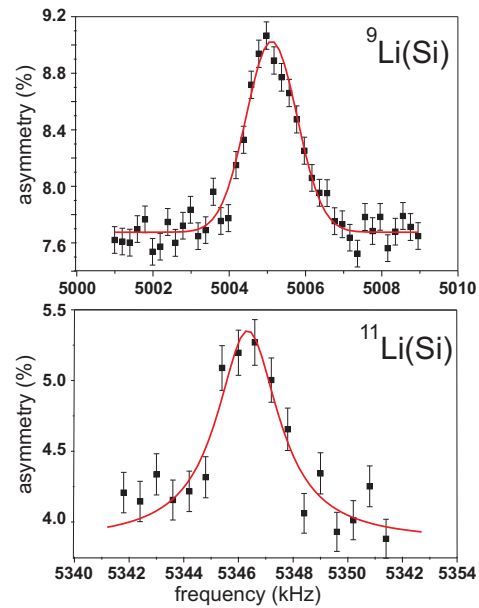}}}
\caption{NMR resonances for $^9$Li and $^{11}$Li in a Si crystal.}
\label{Larmor}
\end{figure}

The results include the magnetic moment $\mu(^9$Li$) =
3.43678(6)\,\mu_N$ which was measured relative to that of $^8$Li.
Now we present a similarly accurate measurement for $^{11}$Li
relative to $^9$Li. Typical NMR resonances of both isotopes
implanted in a Si crystal at room temperature are shown in
Fig.\,\ref{Larmor}. Seven spectra on $^9$Li and three spectra on
$^{11}$Li were recorded in total, while the magnetic field of
about 0.29 T was kept constant. During the measurements, the
magnetic field drifted by 0.005$\%$ at most, which
is consistent with the
scatter of the $^9$Li resonance frequencies and gives an upper
limit for a systematic difference between the fields applied for
both isotopes. The
weighted means of the Larmor
frequencies yield $g(^9$Li$)/g(^{11}$Li$) = 0.93615(6)$ and with
the $g$ factor of $^9$Li \cite{Bor05} we have $g(^{11}$Li$) =
2.44746(17)$. With the spin $I=3/2$ this gives the magnetic moment
$\mu(^{11}$Li$)=3.6712(3)\,\mu_N$, improved by an order of
magnitude and meeting the requirement for a precise measurement of
the quadrupole moment.

The quadrupole moments of $^{8}$Li and $^{9}$Li have been reported
from measurements in several different crystals (see
\cite{Bor05}). Although some of the values seem to be in poor
agreement with one another, we found that for $^8$Li they become
consistent \cite{Bor05}, if they are evaluated with respect to the
same reference value $Q(^7$Li$)= -40.0(3)$ mb. From this we
adopted a weighted mean value of $Q(^8$Li$)= 31.4(2)$ mb. For
$^9$Li, in order to remove a discrepancy between two earlier
measurements, we remeasured the quadrupole moment relative to that
of $^8$Li. Two implantation hosts, a metallic Zn crystal (hcp
structure) and a LiTaO$_3$ crystal (orthorhombic structure) gave
consistent results, leading to $Q(^9$Li$)= -30.6(2)$ mb
\cite{Bor05}.

\begin{figure}[t]
\centerline{\scalebox{0.65}[0.65]{\includegraphics{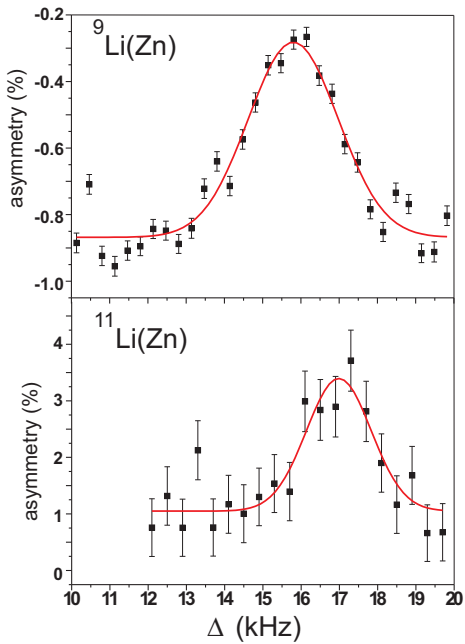}}}
\caption{Multiple-rf resonances for $^9$Li and $^{11}$Li in a Zn
crystal.} \label{NQR}
\end{figure}

\begin{figure}[b]
\vspace*{-0.0cm}
\centerline{\scalebox{0.6}[0.6]{\includegraphics{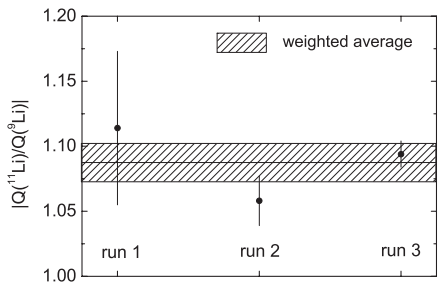}}}
\vspace*{-0.0cm} \caption{The ratio of the $^{11}$Li to $^9$Li
quadrupole moment has been measured in three independent runs. The
weighted average with the error is presented as the dashed bar.}
\label{summary}
\end{figure}

With this precise value,
and by measuring the quad\-ru\-po\-le frequency of $^{11}$Li relative to
that of $^9$Li, we can now determine the $^{11}$Li quadrupole
moment to a similar precision. Three independent experimental runs
were performed, during which the quadrupole splitting $\Delta_r$
was measured successively for $^9$Li and $^{11}$Li implanted in
the Zn crystal. Typical multiple-rf resonances are shown in
Fig.\,\ref{NQR}. Each run started by measuring the Larmor
frequency of $^9$Li in Zn, which was then used to calculate the
Larmor frequency of $^{11}$Li from the $g$-factor ratio given
above. These two values define the center frequencies of the
multiple-rf scans. The
resonance values $\Delta_r$ for
$^{9}$Li and $^{11}$Li directly give the ratio of the quadrupole
moments. Results from the three runs are presented in
Fig.\,\ref{summary}, with the weighted mean of $\vert
Q(^{11}$Li$)/Q(^9$Li$)\vert =1.088(15)$. Thus the quadrupole
moment of the halo nucleus, $\vert Q(^{11}$Li$)\vert=33.3(5)$ mb,
with a negative sign from theoretical considerations, is nearly
10$\%$ larger than that of bare $^9$Li.

\begin{figure}
\centerline{\scalebox{0.4}[0.4]{\includegraphics{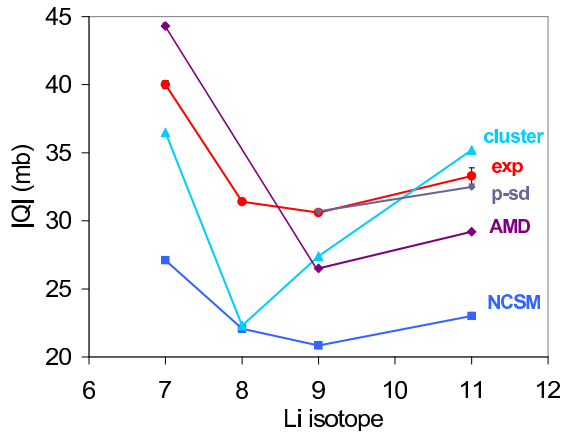}}}
\caption{Quadrupole moments of the Li isotopes compared to values
from different theoretical models.} \label{compare}
\end{figure}

In Table \ref{moments} we summarize the experimental quadrupole
and magnetic moments of the Li isotopes, which are now all known
to about 1$\%$ or better, thus allowing tests of various nuclear
models
(Fig.\,\ref{compare}). In the shell model it is known that some
magic numbers which are valid near stability, disappear in nuclei
with extreme isospin due to the changing strength of the
spin-isospin dependent term in the residual nucleon-nucleon ($NN$)
interaction \cite{Ots01}.  In light neutron-rich nuclei this leads
to the disappearance of the $N=8$ magic number when protons are
taken out of the $\pi p_{3/2}$ orbital. A first experimental proof
for this was found in the positive parity of the $^{11}$Be ground
state, shown to be dominated by $\nu 2s_{1/2}$ \cite{Tal60,Gei99}
and not $\nu 1p_{1/2}$ as expected for
7 neutrons.
In $^{11}$Li this is manifest
in the halo structure which
has about 45$\%$ of $(2s_{1/2})^2$ occupancy \cite{Sim99}.
It is also reflected in the quadrupole moments, but the increase
from $^9$Li to $^{11}$Li can not be understood by an excitation of
neutrons to the spherical $2s$ orbit only.
Comparing our result to large-scale shell model calculations may
provide a clue for some $1d$ occupancy of the halo neutrons.
Suzuki et al.\
modified an effective
shell-model interaction in order to describe
both stable and exotic nuclei in this $p$-$sd$ region assuming
$^4$He as an inert core \cite{Suz03}.
The calculated quadrupole moment of $^{11}$Li relative to that of
$^9$Li approaches the experimental value if the model space is
extended from the $p$ shell only ($Q_p = 28.2$ mb) to include
excitations into the $sd$ shell ($Q_{psd} = 32.5$ mb), and
effective charges $e_p = 1.2e$ and $e_n = 0.3e$ are used
\cite{Suz07}. The halo wave function then has
$50\%$ of $(2s_{1/2})^2$, $25\%$ of $(1p_{1/2})^2$ and $25\%$ of
$(1d_{5/2})^2$ occupation probability.
Advanced Quantum Monte Carlo calculations, including additionally
a three-nucleon potential \cite{Pie01}, have been successful in
describing the moments of $^{7,8,9}$Li \cite{Pie02}, but so far no
results are available for $^{11}$Li.

\begin{table}[b]
\caption{Experimental dipole and quadrupole moments of Li
isotopes from \cite{Bor05}. Results on $^{11}$Li are from this
work, with the sign of $Q(^{11}{\rm Li})$ assumed from theory.
\label{moments} }
 \begin{tabular}{|l|l|l|l|}
 \hline
 isotope    &I$^\pi$   & $\mu(\mu_N)$        & $Q($mb$)$    \\
 \hline
 $^6$Li     &1$^+$         &  0.8220473(6)   & -0.806(6)     \\
 $^7$Li     &3/2$^-$       &  3.256427(2)    & -40.0(3)      \\
 $^8$Li     &2$^+$         &  1.653560(18)   & +31.4(2)      \\
 $^9$Li     &3/2$^-$       &  3.43678(6)     & -30.6(2)      \\
 $^{11}$Li  &3/2$^-$       &  3.6712(3)      & (-)33.3(5)    \\
 \hline
 \end{tabular}
\end{table}

More sophisticated models abandon the assumption of an inert core:
all nucleons are active in the
no-core shell model (NCSM).  Navratil et al.\ have performed such
calculations, using an effective interaction derived
microscopically from a $NN$ potential fitted to $NN$ scattering
data \cite{Nav98}, and they find good agreement with the ground
state properties of $A=7$-11 nuclei. The model does not explain
the charge radii, but it
perfectly reproduces the trend of experimental quadrupole moments
of the Li isotopes. Only an overall scaling factor is missing,
which is explained by a limited model space and might be
compensated by using effective charges.


An alternative approach is to consider the Li nuclei made of
$\alpha$ and triton clusters plus additional neutrons. In the
microscopic cluster model by Varga et al.\ \cite{Var95}, various
cluster arrangements are combined to include all possible
correlations between the clusters.  In this approach, the
quadrupole moments are underestimated, except the one of $^{11}$Li
which is about 30\% larger than that of $^9$Li
\cite{Var02}. This is surprising, because
the model happened to reproduce best the development of the charge
radii \cite{San06}.

The microscopic antisymmetrized molecular dynamics (AMD) approach
\cite{Kan01} has been rather successful in
describing
many features of light nuclei including their electromagnetic
moments. For the quadrupole moments it predicts an 8\% increase
from $^{9}$Li to $^{11}$Li, very close to the experimental value.
However, this agreement should not be overrated as long as halo
properties, in particular the large matter radius of $^{11}$Li,
are not reproduced.

Finally, it is not clear to what extent the different theoretical
approaches have included the center-of-mass motion of the proton
distribution introduced by the two halo neutrons. Therefore we try
to describe the observations assuming the very simple picture of a
$^9$Li core surrounded by two halo neutrons known to be in
spherical orbits. From experiment \cite{San06} we know that the
rms charge radius $\langle r^2 \rangle^{1/2}$ increases by
11(2)\%, from 2.185(33) fm for $^9$Li to 2.423(34) fm for
$^{11}$Li according to the most recent evaluation \cite{Puc06} ,
while we find that the quadrupole moment increases by 8.8(1.5)\%.

If the deformation of the charge distribution would be the same
for $^9$Li and $^{11}$Li, one should expect an increase of the
quadrupole moment proportional to
the mean square
radius $\langle r^2 \rangle$. However, if we ascribe the increase
of the radius to a recoil effect caused by the spherical halo, the
center-of-mass movement produces a spherical expansion of the
(non-spherical) charge distribution for which the quadrupole
moment increases only with the square root $\langle r^2
\rangle^{1/2}$.
This can explain the striking
analogy between
the quadrupole moments
and the rms charge radii without any additional change of the
$^9$Li-core structure caused by the presence of the halo neutrons.
In particular, the $^{11}$Li quadrupole moment seems to be
unaffected by quadrupole core polarization
involving a substantial $1d$ component in the halo wave function.
We
note that this intuitive relationship between the $^{11}$Li
quadrupole moment and charge radius is independent of correlations
in the movement of the halo neutrons, although these affect
strongly the
behavior
of both quantities.

In conclusion, we have measured the quadrupole moment of the
$^{11}$Li halo nucleus relative to that of its $^9$Li core, with a
precision improved by an order of magnitude, thus providing a test
of modern nuclear theories. While these theories have difficulties
to reproduce simultaneously the charge radii and quadrupole
moments of both isotopes, we found a relationship between both
quantities in a simple halo picture that is surprisingly well
fulfilled.

This work has been supported by the German Federal Ministry for
Education and Research (BMBF), contracts no.\ 06 MZ 962 I and 06
MZ 175, the European Union through RII3-EURONS (506065), the IUAP
project P5-07 of OSCT Belgium and the FWO-Vlaanderen. The authors
thank the ISOLDE technical group for their assistance.

\end{document}